\documentclass{PoS}
\title{Exotic Heavy Quark Spectroscopy -- Theory Interpretation vs Data}

\ShortTitle{Exotic Heavy Quark Spectroscopy – Theory Interpretation vs Data}
\author{\speaker{Christian Hambrock}%
        TU Dortmund\\
        E-mail: \email{christian.hambrock@tu-dortmund.de}}

\abstract{
An overview over recent spectroscopy results in comparison with data is given.
We focus on the exotica $Z_b(10610)$, $Z_b(10650)$ and $Y_b(10890)$ in the bottom 
sector and discuss a possible connection to the exotica in the charm sector, namely to the resonances $Y_c(4260)$ and  $Z_c(3900)$. Furthermore, future prospects for the determination of the nature of $Y_b(10890)$ are outlined.
}

\FullConference{14th International Conference on B-Physics at Hadron Machines\\
                 April 8-12, 2013\\
                 Bologna, Italy}

\begin{document}
\section{Introduction}
The term exotica labels states which have an identical number of quarks and antiquarks, but defy an ordinary meson classification. 
Many exotic states in the charm sector with  $c\bar c$ content have been discovered by Belle and others, see~\cite{Zupanc:2009qc,Ablikim:2013mio} and references therein. 
While there are most likely many more which are yet unknown, all of them should also reflect in the $b\bar b$ sector, according to heavy quark symmetry. The non-discovery of the respective $b\bar b$ partners of the charmonium-like exotica would be even more enigmatic. 
Indeed, there exist three candidates up to date, namely the states labeled $Y_b(10890)$, $Z_b(10610)$ and $Z_b(10650)$, observed by Belle~\cite{Belle:2011aa}. 
The scarceness of the discovered  $b\bar b$ states, compared to the charm ones, is mostly owed to the current status of experimental data; the charm sector can be probed by the $e^+e^-$ machines not only directly, but also through $B$ meson decays and initial state radiation. The bottom sector on the other hand is relying on direct measurements, most of which have been tuned to the center-of-mass energy of the masses of $\Upsilon(4S)$ and $\Upsilon(5S)$. The dataset for the latter has roughly one order of magnitude smaller statistics than the former. Notably, all three exotic $b\bar b$ candidates have firstly been discovered in that sample. This is not surprising, since exotica lie generically above the open heavy quark thresholds due to their multiquark nature.
A list of the observed exotica is given in Table~\ref{tab:allstates}.
\begin{table}[b]
	\caption{
		Exotic states found by Belle and others~\cite{Zupanc:2009qc,Ablikim:2013mio}. An asterisks indicates the first discovery before Belle.
}
	\label{tab:allstates}
\center
{\tiny
\begin{tabular}{llllllll}
\hline\hline
State   & $M$~(MeV) & $\Gamma$~(MeV)    & $J^{PC}$ & Decay Modes & 
Production  Modes & Also observed by  & date \\\hline
& & & & &  $e^{+} e^{-} $~(ISR) & &   \\
$Y_s(2175)$ & $2175\pm8$ & $ 58\pm26 $ & 
$1^{--}$ & $\phi f_0(980)$     & $J/\psi\rightarrow\eta Y_s(2175)$ &  BaBar$^*$, BESII & 2006\\
&&&&$\pi^{+}\pi^{-}J/\psi$, & & BaBar &  
\\
$X(3872)$& $3871.4\pm0.6$ & $<2.3$ & $1^{++}$ & $\gamma J/\psi$,$D\bar{D^*}$ & $B\rightarrow KX(3872)$, $p\bar{p}$ & CDF, D0 & 2003
\\
$Z(3900)$& $3899 \pm 6$ & $46\pm 22$ & $1^{+}$ & $\pi^{\pm} J/\psi$ & $Z(4260)\to Z(3900)\pi$ & BESIII$^*$ &2013
\\
$X(3915)$& $3914\pm4$&$ 28^{+12}_{-14} $& $0/2^{++}$ &$\omega J/\psi$ & $\gamma\gamma\rightarrow X(3915)$ & & 2009  \\
$Z(3930)$& $3929\pm5$&$ 29\pm10 $& $2^{++}$ & $D\bar{D}$ & $\gamma\gamma\rightarrow Z(3940)$ &  & 2009 \\
 &&&& $D\bar{D^*}$ (not $D\bar{D}$ && & \\
$X(3940)$ & $3942\pm9$ & $ 37\pm17 $ & $0^{?+}$ &  or $\omega J/\psi$)  & $e^{+} e^{-} \rightarrow J/\psi  X(3940)$ & & 2005  \\
$Y(3940)$& $3943\pm17$&$ 87\pm34 $&$?^{?+}$ & $\omega J/\psi$ (not
$D\bar{D^*}$) & $B\rightarrow K Y(3940)$ &  BaBar & 2005 \\
$Y(4008)$& $4008^{+82}_{-49}$&$ 226^{+97}_{-80}$ &$1^{--}$& $\pi^{+}\pi^{-} J/\psi$ &
$e^{+} e^{-} $(ISR) &  &2005\\
$X(4160)$& $4156\pm29$&$ 139^{+113}_{-65}$ &$0^{?+}$& $D^*\bar{D^*}$
 (not $D\bar{D}$) & $e^{+} e^{-}  \rightarrow J/\psi X(4160)$  &  & 2008\\
$Y(4260)$& $4264\pm12$&$ 83\pm22$ &$1^{--}$&  $\pi^{+}\pi^{-} J/\psi$ & $e^{+} e^{-} $(ISR)
&BaBar$^*$, CLEO  & 2005   \\
$Y(4350)$& $4361\pm13$&$ 74\pm18$ &$1^{--}$&  $\pi^{+}\pi^{-} \psi '$ & $e^{+} e^{-} $(ISR)
& BaBar$^*$ & 2007  \\
$X(4630)$& $4634^{+9}_{-11}$&$ 92^{+41}_{-32} $ &$1^{--}$&  $\Lambda_c^+\Lambda_c^-$ & $e^{+} e^{-}  $(ISR)
&     &2008  \\
$Y(4660)$& $4664\pm12$&$ 48\pm15 $ &$1^{--}$&  $\pi^{+}\pi^{-} \psi '$ & $e^{+} e^{-}  $(ISR)
&    & 2007   \\
$Z(4050)$& $4051^{+24}_{-23}$&$ 82^{+51}_{-29}$ & ? &
$\pi^{\pm}\chi_{c1}$ & $B\rightarrow K Z^{\pm}(4050)$  & &2008  \\
$Z(4250)$& $4248^{+185}_{-45}$&$ 177^{+320}_{-72}$ & ? &
$\pi^{\pm}\chi_{c1}$ & $B\rightarrow K
Z^{\pm}(4250)$  &  & 2008 \\         
$Z(4430)$& $4433\pm 5$&$ 45^{+35}_{-18}$ & ? & $\pi^{\pm}\psi '$ & $B\rightarrow K
Z^{\pm}(4430)$  &  & 2007\\\hline\hline
$Z_b(10610)$& $10,607\pm 2$&$ 18.4 \pm 2.4$ & $1^+$ & $\pi^{\pm} h_b(1,2 P), \pi^{\pm} \Upsilon(1,2,3S)$ & 
$Y_b/\Upsilon(5S) \to Z_b(10610)\pi$  & &2011 \\
$Z_b(10650)$  & $10,652\pm 2$ & $ 11.5 \pm 2.2$ &   $1^+$ &
$\pi^{\pm} h_b(1,2 P), \pi^{\pm} \Upsilon(1,2,3S)$
& $Y_b/\Upsilon(5S) \to Z_b(10650)\pi$ & & 2011 \\
$Y_b(10890)$  & $10,890\pm 3$ & $55\pm 9$ &   $1^{--}$ &
$\pi^{+}\pi^{-}\Upsilon(1,2,3S)$
& $e^{+} e^{-} \rightarrow Y_b$ & & 2008\\
\hline\hline
\end{tabular}%
}
\end{table}

\begin{figure}[t]
\centering
\includegraphics[width=0.9\textwidth]{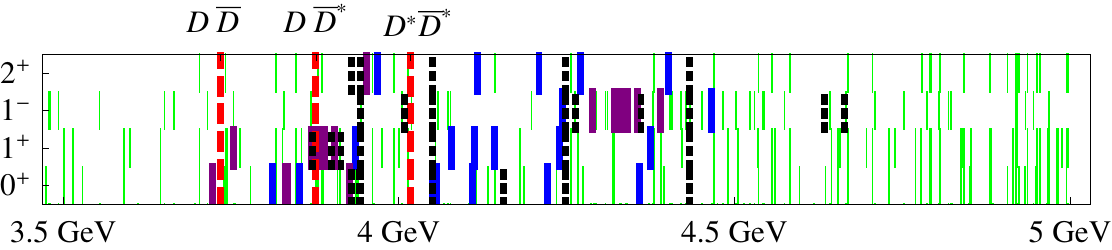} 
\caption{
Status of current mass estimates in the charm sector 
%(see also Tab.~\ref{tab:allstates}) 
and comparison with data.
The data (dashed, black) is shown  with corresponding quantum numbers (if unknown, the line stretches over the full range).
The theory estimates are given for the molecules as an illustrative naive combination of mesons (thin, green lines), where the masses are taken from the PDG~\cite{Beringer:1900zz}.
The tetraquark estimates are taken from~\cite{Ebert:2005nc,Ebert:2007rn} based on the relativistic quark model (blue) and from~\cite{Maiani:2004vq, Drenska:2009cd} based on the constituent quark model (purple).
\label{fig:ccbarstat}
}
\end{figure}

Models to accommodate the exotica have been proposed over the last decades. The molecular interpretation (dimeson states bound similar to nuclei)~\cite{Tornqvist:2004qy,Liu:2005ay,Rosner:2007mu} is favored for some states, while the tetraquark interpretation (four quark states which are genuinely bound by gluons)~\cite{Maiani:2004vq,Maiani:2005pe,Drenska:2008gr} is favored for others. Hybrids~\cite{Kou:2005gt} (quarks plus glueballs) play a slightly minor role in the current discussion, but are also studied in this context.  However, a consistent picture is missing. None of the present models can explain all states simultaneously, while in addition more predicted states are missing than are discovered in either ansatz. 
The status in the charm sector is sketched in Figure~\ref{fig:ccbarstat}.
For the molecular interpretation, it is not clear, which combination of mesons should bind; there is currently no deeper understanding, why some states are observed (like for example the $X_c(3872)$, which is a candidate for a $D^0 \bar D^0$ bound state), while a plethora of different possible combinations is absent. On the other hand, the tetraquark picture is incomplete, since reliable calculations for the masses are missing.
The calculations in the relativistic quark model~\cite{Ebert:2005nc, Ebert:2007rn} are not in good agreement with  constituent model estimates \cite{Drenska:2009cd,Maiani:2004vq}. In addition, neither of the two approaches have good overall agreement with the experiment.
%%\cite{Drenska:2009cd}
%\bibitem{Drenska:2009cd} 
%  N.~V.~Drenska, R.~Faccini, A.~D.~Polosa and ,
%  %``Exotic Hadrons with Hidden Charm and Strangeness,''
%  Phys.\ Rev.\ D {\bf 79}, 077502 (2009)
%  [arXiv:0902.2803 [hep-ph]].
%  %%CITATION = ARXIV:0902.2803;%%
%  %41 citations counted in INSPIRE as of 03 Apr 2013
%  %\cite{Maiani:2004vq}
%  \bibitem{Maiani:2004vq} 
%    L.~Maiani, F.~Piccinini, A.~D.~Polosa, V.~Riquer and ,
%    %``Diquark-antidiquarks with hidden or open charm and the nature of X(3872),''
%    Phys.\ Rev.\ D {\bf 71}, 014028 (2005)
%    [hep-ph/0412098].
%    %%CITATION = HEP-PH/0412098;%%
%    %305 citations counted in INSPIRE as of 03 Apr 2013
%%\cite{Ebert:2005nc}
%\bibitem{Ebert:2005nc} 
%  D.~Ebert, R.~N.~Faustov, V.~O.~Galkin and ,
%  %``Masses of heavy tetraquarks in the relativistic quark model,''
%  Phys.\ Lett.\ B {\bf 634}, 214 (2006)
%  [hep-ph/0512230].
%  %%CITATION = HEP-PH/0512230;%%
%  %86 citations counted in INSPIRE as of 03 Apr 2013
%%\cite{Ebert:2007rn}
%\bibitem{Ebert:2007rn} 
%  D.~Ebert, R.~N.~Faustov, V.~O.~Galkin, W.~Lucha and ,
%  %``Masses of tetraquarks with two heavy quarks in the relativistic quark model,''
%  Phys.\ Rev.\ D {\bf 76}, 114015 (2007)
%  [arXiv:0706.3853 [hep-ph]].
%  %%CITATION = ARXIV:0706.3853;%%
%  %14 citations counted in INSPIRE as of 03 Apr 2013

For  a theoretical interpretation of the data it is insufficient to rely on the mass estimates only. Hopefully the lattice will provide unbiased input for the masses and fuel the discussion eventually. Phenomenologically, however, one needs to study further characteristics, such as decay patterns. But 
it is also imperative to find relations among the different states to test an underlying model on a broad basis.
The heavy quark symmetry should hold to some extent. 
We discuss one important puzzle and point out a possible relation between patterns in both heavy quark sectors. It is argued that the states $Y_c(4260)$ and $Y_b(10890)$, as well as the states $Z_c(3900)$ and $Z_b(10610,10650)$, might be heavy quark partners.

\section{$Y_b(10890)$}\label{sec:Yb}
The potential exotic state $Y_b(10890)$ with $J^{PC}=1^{--}$ was first observed by the Belle collaboration~\cite{Abe:2007tk,Adachi:2008pu} and remains to be confirmed by independent experiments.
The anomalously large production cross sections for $e^+e^- \to \Upsilon(1S,2S,3S) \pi^+ \pi^- $ measured at $\Upsilon(5S)$  did not agree well with
the lineshape and production rates for the conventional $b\bar{b}$ state $\Upsilon(5S)$. 
Ordinary $\Upsilon(nS)$ decays are well-described by the multipole expansion~\cite{Brown:1975dz}, which involves coupling to two gluons and are hence Zweig forbidden and do not show any significant resonant structure. 
Compared to the ordinary $\Upsilon(nS) \to \Upsilon(mS) \pi^+ \pi^-, m<n$ decays, the derived partial width at $\Upsilon(5S)$ is out of line by two orders of magnitude. In addition, a distinct resonant structure is observed in the invariant mass spectra. 
Currently two theoretical explanations are competing; the tetraquark interpretation on the one hand~\cite{Ali:2009es,Ali:2010pq} and the rescattering model~\cite{Meng:2007tk}  on the other. The former can explain the enhancement and the resonant structure via Zweig allowed decays and coupling to intermediate resonances, while the latter is relying on the decay $\Upsilon(5S)\to B^{(*)}\bar B^{(*)}$ and a subsequent recombination of the $B$ mesons.

\section{$Z_b(10610)$ and $Z_b(10650)$}

Belle~\cite{Belle:2011aa} reported the measurement of the
 $\pi^\pm \Upsilon(nS)(n=1,2,3)$ and $\pi^\pm h_b(mP) (m=1,2)$ invariant mass spectra
from the data taken  near the peak of  the $\Upsilon(5S)$ resonance in the processes $e^+e^-\to \Upsilon(nS) \pi^+\pi^-$
 and $ e^+e^-\to h_b(mP)\pi^+\pi^-$, in which two charged bottomonium-like states
 $Z^\pm_b(10610)$ and $Z^\pm_b(10650)$ were discovered. It is not clear if the $Z_b$ states are the decay products of $Y_b(10890)$ or of $\Upsilon(5S)$, as the Belle measurements are not equivocal on this point, and the production mechanism, obviously central to theoretical interpretation, remains to be unambiguously  confirmed.   
 The angular distribution analysis indicates that the 
 quantum numbers of both $Z^\pm_b$ and $Z'^\pm_b$ are 
 $I^{G}(J^P)=1^+(1^+)$, where $ Z_b$ and $ Z_b'$ are the lighter and heavier flavor eigenstates, which are identified for small mixing with $Z_b(10610)$ and $Z_b(10650)$, respectively. Their neutral isospin counterparts with $I_3=0$ have 
$J^{PC}=1^{+-}$.
The masses and decay widths averaged over the five different final states
 are~\cite{Belle:2011aa}:
 \begin{equation}
\begin{array}{cccccc}
m_{Z^\pm_b}&=&10607.2\pm2.0\,  \rm MeV
,&
m_{Z'^\pm_b}&=&10652.2\pm1.5\,   \rm MeV
,\nonumber\\
\Gamma_{Z^\pm_b}&=&18.4\pm2.4\,   \rm MeV
,&
\Gamma_{Z'^\pm_b}&=&11.5\pm2.2\,  \rm MeV
.\label{eq:belledata}
\end{array}
 \end{equation} 
Due to the proximity of the  $Z_b$ and $Z_b^\prime$ masses with the $B\bar{B}^*$ and $B^*\bar{B}^*$ thresholds~\cite{Nakamura:2010zzi}, it has been proposed that the former   could be realized as $S$-wave $B\bar{B}^*$ and 
$B^*\bar{B}^*$  molecular states, respectively~\cite{Bondar:2011ev,Voloshin:2011qa,Zhang:2011jja,Yang:2011rp,Sun:2011uh,Cleven:2011gp,Mehen:2011yh}. 
In this scenario, the heavy quark spin
structure of the $Z_b$ and $Z_b^\prime$ is expected to mimic that of the corresponding
meson pairs, which is given  in a non-relativistic notation by
\begin{eqnarray}
| Z_b\rangle
&=& (0^-_{b\bar q}\otimes 1^-_{q\bar b} +1^-_{q\bar b}\otimes 0^-_{q\bar b})/\sqrt 2,\nonumber\\
 | Z_b^\prime\rangle
&=&  1^-_{b\bar q}\otimes 1^-_{q\bar b} ,
 \label{eq:Fierz-Bondar}
\end{eqnarray} 
where $0^-$ and $1^-$ denote the para and ortho states with negative parity, respectively. 
One anticipates 
the mass splitting to follow $\Delta m_{Z_b} \equiv  m_{Z_b^\prime}-m_{Z_b}= m_{B^*}-m_{B} \simeq 46$ MeV,
in neat agreement with the observed value $\Delta m_{Z_b} = (45\pm 2.5)$ MeV~\cite{Belle:2011aa}. Moreover,
the structure in Eq.~(\ref{eq:Fierz-Bondar})  predicts that  $Z_b$ and $Z_b^\prime$  should have 
the same decay width, which is  approximately in agreement with the data.

There is a small caveat, since the masses of the states lie above their respective thresholds by about 2 MeV. 
If consolidated by more precise experiments, this feature may become a
serious problem in this approach, as a one-pion exchange potential, which would
 produce such a bound state, does not support an $S$-wave
$B \bar{B}^*$ resonance above threshold in an effective field theory~\cite{Nieves:2011vw}.
Also, the measured total decay widths appear much too large compared to the 
naively expected ones for loosely bound states, and this suggests that both $Z_b$ states
are compact hadrons. 

In our previous work~\cite{Ali:2009pi}, we calculated the $Z_b$ masses in the tetraquark model based on an  effective Hamiltonian
approach. The agreement with the experimental masses is not particularly good. However, we showed in~\cite{Ali:2011ug} that by including meson loop contributions and mixing between the heavier and the lighter $Z_b$ states, the measured masses can in principle be obtained in parts of the parameter space (depending on the coupling to the dimeson channels).

A more striking pattern emerges  based on the heavy quark spin symmetry. There is an analogous expression to the decomposition in~(\ref{eq:Fierz-Bondar}) for the tetraquark states
\begin{eqnarray}
| Z_b\rangle&=& \big(0_{[bq]}\otimes 1_{[\bar{b}\bar{q}]} -1_{[bq]}\otimes 0_{[\bar{b}\bar{q}]}\big)/\sqrt 2,\nonumber\\ 
| Z_b^\prime\rangle&=& 1_{[bq]}\otimes 1_{ [\bar b\bar q]},
\end{eqnarray}
where square brackets denote the composite diquarks, 
 see~\cite{Ali:2009pi} for details. 
Performing a Fierz transformation, the flavor and spin content in the $b\bar q\otimes q\bar b$ product space can be made explicit:
\begin{eqnarray}
| Z_b\rangle
&=& 1^-_{b\bar q}\otimes 1^-_{q\bar b},\nonumber\\
| Z_b^\prime\rangle 
&=& ( 1^-_{b\bar q}\otimes 0^-_{q\bar b}+0^-_{b\bar q}\otimes 1^-_{q\bar b})/\sqrt 2.
\label{eq:Fierz}
\end{eqnarray} 
The labels $0_{b\bar q}$ and $1_{b\bar q}$ 
in  Eq.~(\ref{eq:Fierz}) can be viewed as $\bar B$ and $\bar B^*$,
respectively. Eq.~(\ref{eq:Fierz}) shows that the $ Z_b$ and $ Z_b'$  have
similar coupling strengths with different final states.
It follows that
$ Z_b$  couples to $B^*\bar B^*$
state  while $ Z_b^\prime$ couples to $B\bar B^*$.

After our prediction of this peculiar decay pattern, the Belle collaboration published new measurements on the $Z_b$ decays~\cite{fortheBelle:2013vwa}, shown in Table~\ref{tab:zb_dec}.
The Belle observation, if confirmed by independent experiments, would
suggest a dominant molecular component. 
\begin{table}[t]
	\caption{Branching fractions ($\mathcal B$) of $Z_b(10610)$ and $Z_b(10650)$ assuming
	that the observed so far channels saturate their decays (Table from~\cite{fortheBelle:2013vwa}). }
\label{tab:zb_dec}
\center
\begin{tabular}{lcc}
\hline\hline
Channel & $\mathcal B$  $(Z_b(10610))$[\%] & $\mathcal B$ of $(Z_b(10650))$[\%] \\
\hline\hline
$\Upsilon(1S)\pi^+ $  & $0.32\pm0.09$ & $0.24\pm0.07$ \\
$\Upsilon(2S)\pi^+$  & $4.38\pm1.21$ & $2.40\pm0.63$ \\
$\Upsilon(3S)\pi^+$ & $2.15\pm0.56$ & $1.64\pm0.40$ \\
$h_b(1P)\pi^+$  & $2.81\pm1.10$ & $7.43\pm2.70$ \\
$h_b(2P)\pi^+$ & $2.15\pm0.56$ & $14.8\pm6.22$ \\
$B^+\bar{B}^{*0}+\bar{B}^0B^{*+}$ & $86.0\pm3.6$ & -- \\
$B^{*+}\bar{B}^{*0}$ & -- & $73.4\pm7.0$ \\
\hline\hline
\end{tabular}
\end{table}

\section{$Y_c(4260)$}
The state $Y_c(4260)$ was first observed by BaBar~\cite{Aubert:2005rm} in the final state $J/\psi \pi^+ \pi^-$. The direct production in $e^+e^-$ annihilation indicates $J^{PC}=1^{--}$ quantum numbers.
There are some striking similarities between the two exotic states $Y_b(10890)$ and $Y_c(4260)$. 
They  not only have the same quantum numbers, but are furthermore observed in the identical process $H \pi^+ \pi^-$, in which $H$ labels a quarkonium $Q\bar Q$  state with $J^{PC}=1^{--}$, with $Q$ being either a $c$ or a $b$ quark (namely $H$ is a member of the $J/\psi$ or $\Upsilon$ family).
A naive mass estimate, starting from our constituent quark model estimate for $Y_b(10890)$~\cite{Ali:2009pi} is obtained by
 subtracting twice the difference of the constituent quark masses $m_b-m_c\approx 3333$~MeV:
\begin{equation}
m_{Y_b(10890)} \approx m_{Y_c(4260)} +2 (m_b - m_c).
\end{equation}
The estimate is only off by about $30$~MeV. However, this feature has to be taken with some caution, since the use of heavy quark symmetry in providing reliable mass estimates
in multiquark systems remains to be quantitatively tested.

Due to the apparent similarities between $Y_c(4260)$ and $Y_b(10890)$, we suggested to search for the partners of the $Z_b$ states in the $Y(4260)$ decays via the channel $Y(4260) \to Z_c^\pm \pi^\mp$~\cite{Ali:2009pi}. 

\section{$Z_c(3900)$}
Indeed in the proposed decay, BESIII observed a state with identical quantum numbers compared to the $Z_b$ states in spring 2013~\cite{Ablikim:2013mio}, namely the $Z_c(3900)$. 
Several aspects are, however, peculiar in that observation. Only one state was found, of which the mass is moreover roughly $4 \sigma$ above the corresponding $D\bar D^{(*)}$ threshold; a finding which is hard to explain in terms of hadronic molecules. 
But there are some aspects which facilitate the tetraquark interpretation~\cite{Faccini:2013lda}.

\section{Prospects for the bottom sector}\label{sec:prop}
\begin{figure}[b]
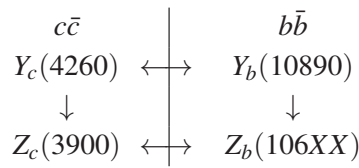

\centering
\begin{tabular}{cc|cc}
	$c\bar c$ && & $b\bar b$
\\
$Y_c(4260)$ && $\hspace{-6mm}\longleftrightarrow$ & $Y_b(10890)$
\\
$\downarrow$&&&$\downarrow$
\\
$Z_c(3900)$ && $\hspace{-6mm}\longleftrightarrow$ & $Z_b(106XX)$
\end{tabular}
\caption{\label{figpuzzle}
	Exotica: Puzzle in the heavy quarks sectors.
}
\end{figure}

Currently there are pending, unanswered questions concerning the exotic spectroscopy in the heavy quark sectors. One big puzzle, depicted in Figure~\ref{figpuzzle}, poses an intriguing mystery. Are the exotic states related? If yes, why do the $Z_c$ and $Z_b$ states appear to be very different? If not, what is the true relation linking the two heavy quark sectors?

To promote the endeavor of understanding the heavy exotic states, the exploration of the bottom sector is important.
Not only new states are waiting to be discovered, but also the existence of $Y_b(10890)$ needs to be established or refuted.
 
\begin{figure}[t]
\includegraphics[width=0.9\textwidth]{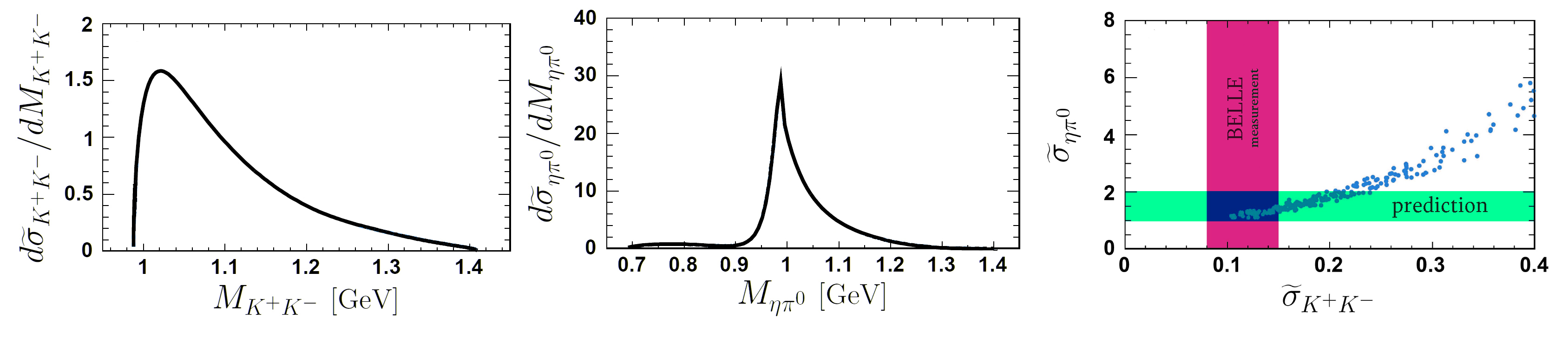} 
\caption{From left to right: Predictions  of the $M_{K^+K^-}$ distribution, of the $M_{\eta \pi^0}$ distribution and  of the correlation between the cross
  sections of $\sigma_{\Upsilon(1S) K^+K^-}$ and $\sigma_{\Upsilon(1S) \eta \pi^0}$, normalized by the measured cross
  section for the $\sigma_{\Upsilon(1S) \pi^+\pi^-}$ mode. In the two left figures, the dotted (solid)
  curves show the dimeson invariant mass spectra from the resonant
  (total) contribution. In the right figure, the dots represent predictions
  from our fit solutions satisfying $\chi^2/{\rm{d.o.f.}}< 1.6$. 
  The shaded  bands shows the current Belle measurement 
  $\widetilde{\sigma}_{K^+K^-}= 0.11^{+0.04}_{-0.03}$~\cite{Abe:2007tk} and the prediction for $\sigma_{\Upsilon(1S) \eta \pi^0}$ (the tilde indicates normalization to $\sigma_{\Upsilon(1S) \pi^+\pi^-}$).  
\label{fig:predictions}
}
\end{figure}

To achieve this intermediate goal, several opportunities arise, which are outlined in the following, one of which has been proposed in~\cite{Ali:2010pq}. Based on the tetraquark model, we developed the formalism for the
processes 
\begin{equation}
	e^+ + e^-\to Y_b\to \Upsilon(1S) + P + P^\prime\,,
\label{eq:process}
\end{equation}
where $P P'$ stands for the pseudoscalar-meson pairs $\pi^+\pi^-$, $K^+K^-$ and
$\eta\pi^0$.  
With this formalism, we analyzed the invariant-mass $M_{PP'}$ and the
$\cos \theta$ spectra, where $\theta$ is the
angle between the momenta of $Y_b$ and $P$ in the $PP'$ rest frame. 
The resulting correlations among
$\sigma_{\Upsilon(1S) \pi^+\pi^-}$, $\sigma_{\Upsilon(1S) K^+K^-}$ and
$\sigma_{\Upsilon(1S) \eta \pi^0}$ are worked out. Constraining these
correlations using the existing data on the first two processes, we
predict $\sigma_{\Upsilon(1S) \eta \pi^0}/\sigma_{\Upsilon(1S) \pi^+\pi^-}$. We also predict
$\sigma_{\Upsilon(1S) K^+K^-}/\sigma_{\Upsilon(1S) K^0 \bar{K}^0}=1/4$, reflecting the
ratio $Q_{[bu]}^2/Q_{[bd]}^2$  with $Q_{[bu]}=1/3$ and $Q_{[bd]}=-2/3$
being the effective electric charges for the constituent diquarks of
the flavor eigenstates. The predicted spectra in Figure~\ref{fig:predictions} show a distinct resonant behavior and are clearly distinguishable from models like the multipole expansion.
We hope, that Belle can improve their current analysis and provide sufficient statistics to show the predicted spectra.

Another way to scrutinize $Y_b(10890)$ is via hadroproduction. Finally, the bottom exotica should show in hadronic processes for which $X_c(3872)$, measured by CDF~\cite{Khachatryan:2010zg}, is a good example. 
To distinguish the rescattering model from the tetraquark explanation, discussed in Sec.~\ref{sec:Yb}, the decay characteristics of $\Upsilon(6S)$ need to be measured.
In the former, the partial width for the $\Upsilon(5S,6S)\to\Upsilon(1S,2S,3S)\pi^+\pi^-$ decays are predicted to be roughly of the same order~\cite{Meng:2007tk}.
The (non-)observation of such a pattern will finally refute (or confirm) one of the few exotic candidates in the bottom sector.
The theoretical work, based on NRQCD, to enable the experimental analysis is currently in preparation~\cite{AhmedWeiI}.

\section{Conclusion}
The theoretical interpretation to the exotic spectroscopy is inconclusive and a big picture to accommodate all exotic states is missing.
A simultaneous explanation of more than just a few states poses a hurdle for all models currently discussed. 
However, eventually consistent patterns need to emerge and be explained, of which one, pictured in Figure~\ref{figpuzzle}, was discussed here.
The completion of the list of bottom exotica, but also a confirmation of $Y_b(10890)$, discussed in Sec.~\ref{sec:prop}, will be a major step in the direction of testing the models and provide theorists with vital input to present a credible explanation of this new form of QCD.

\end{document}